# Linear and nonlinear optical absorption of position-dependent mass oscillators

J. P. G. Nascimento* and I. Guedes

Departamento de Física, Universidade Federal do Ceará, Campus do PICI, Caixa Postal 6030, 60455-760, Fortaleza, CE, Brazil.

**Abstract**

We study the linear $(\alpha^{(1)})$, nonlinear $(\alpha^{(3)})$ and total $(\alpha)$ optical absorptions of position-dependent mass oscillators (PDMOs). We consider three mass distributions $(m(x,\lambda))$ used to describe semiconducting structures; $\lambda$ is a deformation parameter. In the limit $\lambda \to 0$, the three systems describe electrons in a parabolic quantum well. For the system $m_1(x) = \frac{m_0}{[1+(\lambda x)^2]^2}$ we observe that $\alpha^{(1)}(\omega)$ $\left(\alpha^{(3)}(\omega)\right)$ increases (decreases) with increasing $\lambda$. For $m_2(x) = m_0[1 + (\lambda x)^2]$ and $m_3(x) = m_0[1 + \tanh^2(\lambda x)]$ the opposite occurs. In the light of the PDMO approach we observe the $m_2(x)$ and $m_3(x)$ systems are very similar, and can not be distinguished by optical transitions between the two lowest electronic levels. We also discussed about the total optical absorption of the systems.

*Corresponding author. Tel.: +55 85997797665.

E-mail address: joaopedro@fisica.ufc.br (J. P. G. Nascimento)

# 1. Introduction

The nonlinear optical properties of quantum dots (QDs) and quantum wells (QWs) in semiconductors, described by position-dependent mass (PDM) Hamiltonians, were calculated [1-9]. The PDM Hamiltonians introduced by von Hoos [10] have the general expression

$$H = \frac{1}{4}\left[m^\gamma(r)\, p\, m^\beta(r)\, p\, m^\alpha(r) + m^\alpha(r)\, p\, m^\beta(r)\, p\, m^\gamma(r)\right] + V(r), \quad (1.1)$$

where $\alpha, \beta, \gamma \in \mathbb{R}$ and $\alpha + \beta + \gamma = -1$. According to Refs. [11-13], to obtain the correct continuity conditions at the abrupt heterojunction between two crystals the condition $\alpha = \gamma$ has to be fullfiled. The BenDaniel-Duke (BDD) Hamiltonian, where $\alpha = \gamma = 0$ and $\beta = -1$, satisfies this condition and therefore has been considered to describe QW systems.

Some authors have recently reported on the calculation of the optical properties of systems described by the BDD Hamiltonian [1-9]. In 2016, Li et al [14] considered a mass distribution $m(x) = m_0\, e^{\lambda x}$ and a confining external potential, $V(x) = (Ce^{\lambda x}/\lambda^2) - (D\lambda^2/e^{\lambda x})$, to calculate the third-harmonic generation (THG) coefficients, the optical absorption (OA) coefficients and the refractive index changes (RIC). In Ref. [15], Hu et al considered $m(x) = m_0 \text{sech}^2(\lambda x)$ and calculated the OA coeffficients and RIC.

Another PDM Hamiltonian satisfying the condition $\alpha = \gamma$ is the position-dependent mass oscillator (PDMO) Hamiltonian, where $\alpha = \gamma = -\frac{1}{4}$ and $\beta = -1/2$. Here we consider a system described by the PDMO Hamiltonian and calculate the linear and nonlinear OA coefficients for three different mass distributions, namely: (i) $m_1(x) = \frac{m_0}{[1+(\lambda x)^2]^2}$, which is similar to $m(x) = m_0 \text{sech}^2(\lambda x)$ used in [16-18]; (ii) $m_2(x) = m_0[1 + (\lambda x)^2]$, which may be useful to analyze the $GaAs/Al_xGa_{1-x}As$ structures [13]; and (iii) $m_3(x) = m_0[1 + \tanh^2(\lambda x)]$, which is analogue to $m(x) = m_a + (1 -$ $e^{-(x/a)^2}$) used to study confined electronic states in a Gaussian graded $Al_y Ga_{1-y} As$ structure [19].

In Section 2, we briefly describe the basic equations to calculate the linear and nonlinear OA coefficients for PDMO systems. In Section 3, we numerically calculate the OA coefficients for PDMO with $m_1(x)$, $m_2(x)$ and $m_3(x)$. In Section 4, we summarize the results.

**2. Basic equations.**

The complete analysis used to obtain the PDMO Hamiltonian is presented in Refs. [20, 21] and it can be written as

$$H(x,p) = -\frac{1}{2}\frac{1}{\sqrt[4]{m(x)}}\frac{d}{dx}\frac{1}{\sqrt{m(x)}}\frac{d}{dx}\frac{1}{\sqrt[4]{m(x)}} + \frac{1}{2}\left(\int^x \sqrt{m(y)}\,dy + X_0\right)^2, \quad (2.1)$$

whose eigenvalue equation reads

$$H(x,p)\,\psi_n(x) = E_n \psi_n(x), \qquad (2.2)$$

with $E_n = n + 1/2$. To find $\psi_n(x)$ we perform the following point canonical transformation

$$X = X(x) = \int^x \sqrt{m(y)}\,dy, \qquad (2.3)$$

$$\psi_n(x) = \sqrt[4]{m(x)}\,\varphi_n[X(x)], \qquad (2.4)$$

into Eq. (2.2) to obtain

$$-\frac{1}{2}\frac{d^2\varphi_n(X)}{dX^2} + \frac{1}{2}(X+X_0)^2\,\varphi_n(X) = E_n\varphi_n(X), \tag{2.5}$$

which is the eigenvalue equation of the constant mass harmonic oscillator displaced by $-X_0$ whose solutions are well known. Thus, the eigenfunctions $\psi_n(x)$ read

$$\psi_n(x) = \frac{1}{\sqrt{\pi^{1/2}\,2^n\,n!}}\sqrt[4]{m(x)}\,exp\left[-\frac{1}{2}\left(\int^x \sqrt{m(y)}\,dy + X_0\right)^2\right]$$
$$\times H_n\left[\int^x \sqrt{m(y)}\,dy + X_0\right], \tag{2.6}$$

and satisfy the normalization condition,

$$\int |\psi_n(x)|^2\,dx = \int |\varphi_n(X)|^2\,dX = 1. \tag{2.7}$$

To obtain the linear and nonlinear OA coefficients, let us consider the compact-density-matrix approach [22]. Suppose that the PDMO system is excited by the incident time-dependent electric field $\boldsymbol{E}(t) = E(t)\hat{\boldsymbol{x}} = E_0 cos(\omega t)\hat{\boldsymbol{x}} = \left[\tilde{E}e^{i\omega t} + \tilde{E}e^{-i\omega t}\right]\hat{\boldsymbol{x}}$, where $\omega$ is the incident photon frequency, $E_0$ is the oscillation amplitude of $E(t)$ and $\tilde{E} = E_0/2$. Let $\rho$ be the density matrix of the system for this regime. The time evolution of the matrix elements of $\rho$ is given by

$$\frac{\partial \rho_{ij}}{\partial t} = \frac{1}{i\hbar}[H_0 - eE(t)x, \rho]_{ij} - \Gamma_{ij}(\rho - \rho^{(0)})_{ij}, \tag{2.8}$$

where $H = H_0 - eE(t)x$ is the Hamiltonian of the system due to its interaction with $\boldsymbol{E}(t)$, $H_0$ (given by Eq. (2.1)) and $\rho^{(0)}$ are, respectively, the Hamiltonian and the density operator of the system for $\boldsymbol{E}(t) = \boldsymbol{0}$, $e$ is the electron charge, and $\Gamma$ is the phenomenological operator responsible for the damping due to the electron-phonon interaction, collisions among electrons, etc. It is assumed that $\Gamma$ is a diagonal matrix and its elements are given by $\Gamma_{ij} = \Gamma_0 = 1/T_0$, where $T_0$ is the relaxation time.

We solve Eq. (2.8) by employing the iterative method [23]

$$\rho(t) = \sum_n \rho^{(n)}(t), \qquad (2.9)$$

where

$$\frac{\partial \rho_{ij}^{(n+1)}}{\partial t} = \frac{1}{i\hbar}\left\{[H_0, \rho^{(n+1)}]_{ij} - i\hbar\Gamma_{ij}\,\rho_{ij}^{(n+1)}\right\} - \frac{1}{i\hbar}[ex, \rho^{(n)}]_{ij} E(t). \quad (2.10)$$

By considering optical transitions only between two-level electronic systems, the electronic polarization $P(t)$ and susceptibility $\chi(t)$ read [22]

$$P(t) = \varepsilon_0 \chi(\omega)\tilde{E}e^{-i\omega t} + \varepsilon_0 \chi(-\omega)\tilde{E}^* e^{i\omega t} = \frac{1}{V}Tr(\rho M), \qquad (2.11)$$

where $M$ is the dipole moment operator, $*$ stands for the complex conjugation, $\varepsilon_0$ is the vacuum permittivity, $V$ is the volume of the system and $Tr$ is the trace of the matrix.

The relation between the susceptibility $\chi(\omega)$ and the optical absorption coefficient $\alpha(\omega)$ is given by

$$\alpha(\omega) = \omega\sqrt{\frac{\mu}{\varepsilon_R}}Im[\varepsilon_0\chi(\omega)], \qquad (2.12)$$

where $\mu$ is the permeability, $\varepsilon_R$ is the real part of the permittivity, and $Im$ stands by the imaginary part of a complex number.

By using Eqs. (2.10)-(2.12), the linear and nonlinear OA coefficients read [22]

$$\alpha^{(1)}(\omega) = \omega \sqrt{\frac{\mu}{\varepsilon_R}} \frac{|M_{10}|^2 \hbar \sigma_v \Gamma_0}{(E_{10} - \hbar\omega)^2 + (\hbar\Gamma_0)^2}, \qquad (2.13)$$

$$\alpha^{(3)}(\omega, I) = -\omega \sqrt{\frac{\mu}{\varepsilon_R}} \left(\frac{I}{2\varepsilon_0 n_r c}\right) \frac{|M_{10}|^2 \hbar \sigma_v \Gamma_0}{[(E_{10} - \hbar\omega)^2 + (\hbar\Gamma_0)^2]^2}$$

$$\times \left\{ 4|M_{10}|^2 - \frac{|M_{11} - M_{00}|^2 [3(E_{10})^2 - 4E_{10}\hbar\omega + \hbar^2(\omega^2 - \Gamma_0^2)]}{(E_{10})^2 + (\hbar\Gamma_0)^2} \right\}, (2.14)$$

where $M_{ij} = \langle \psi_i | ex | \psi_j \rangle$ are the dipole moment matrix elements, $E_{10} = E_1 - E_0$, $\psi_n$ and $E_n$ are the eigenfunctions and eigenvalues of $H_0$, respectively, $\sigma_v$ is the electronic density of the system, $c$ is the speed of light in vacuum, $n_r$ is the refractive index and $I = 2\varepsilon_0 n_r c |\tilde{E}|^2$ is the incident optical intensity.

From Eqs. (2.13) and (2.14), the total OA coefficient is given by

$$\alpha(\omega, I) = \alpha^{(1)}(\omega) + \alpha^{(3)}(\omega, I). \qquad (2.15)$$

3. Results and discussion.

From Eq. (2.6) the eigenfunctions for $m_1(x), m_2(x)$ and $m_3(x)$, are, respectively, given by

$$\psi_{1,n}(x) = \frac{\sqrt[4]{m_0[1 + (\lambda x)^2]^{-2}}}{\sqrt{\pi^{1/2} 2^n n!}} \exp\left\{ -\frac{m_0}{2} \left[ \frac{\arctan(\lambda x)}{\lambda} \right]^2 \right\}$$

$$\times H_n\left\{ \sqrt{m_0} \left[ \frac{\arctan(\lambda x)}{\lambda} \right] \right\}, \qquad (3.1)$$

$$\psi_{2,n}(x) = \frac{\sqrt[4]{m_0[1+(\lambda x)^2]}}{\sqrt{\pi^{1/2}\, 2^n\, n!}} exp\left\{-\frac{m_0}{2}\left[\frac{x}{2}\sqrt{1+(\lambda x)^2}+\frac{\arcsinh(\lambda x)}{2\lambda}\right]^2\right\}$$

$$\times H_n\left\{\sqrt{m_0}\left[\frac{x}{2}\sqrt{1+(\lambda x)^2}+\frac{\arcsinh(\lambda x)}{2\lambda}\right]\right\}, \qquad (3.2)$$

and

$$\psi_{3,n}(x) = \frac{\sqrt[4]{m_0[1+\tanh^2(\lambda x)]}}{\sqrt{\pi^{1/2}\, 2^n\, n!}}$$

$$\times exp\left\{-\frac{m_0}{2}\left[\sqrt{2}\,\arcsinh[\sqrt{2}\sinh(\lambda x)]-\arctanh\left[\frac{\sinh(\lambda x)}{\sqrt{\cosh(2\lambda x)}}\right]\right]^2\right.$$

$$\times \frac{\cosh^2(\lambda x)[1+\tanh^2(\lambda x)]}{\lambda^2 \cosh(2\lambda x)}\right\}$$

$$\times H_n\left\{\sqrt{m_0}\left[\sqrt{2}\,\arcsinh[\sqrt{2}\sinh(\lambda x)]-\arctanh\left[\frac{\sinh(\lambda x)}{\sqrt{\cosh(2\lambda x)}}\right]\right]\right.$$

$$\left.\times \frac{Cosh(\lambda x)\sqrt{[1+\tanh^2(\lambda x)]}}{\lambda\sqrt{Cosh(2\lambda x)}}\right\}. \qquad (3.3)$$

In Figs. 1(a)-(b), we show plots of $\alpha^{(1)}(\omega)$ and $\alpha^{(3)}(\omega)$, respectively, for $m_1(x), m_2(x)$ and $m_3(x)$ with $\lambda = 0.6$. We observe that the peak (minimum) of $\alpha^{(1)}$ $(\alpha^{(3)})$ for the three distributions always occurs at $\omega = 1.4$ ($\omega = 1.2$). The value of $\alpha^{(1)}$ $(\alpha^{(3)})$ of $m_1(x)$ system is higher (lower) than those of $m_2(x)$ and $m_3(x)$ ones. Observe that the spectra of $\alpha^{(1)}$ and $\alpha^{(3)}$ of $m_2(x)$ and $m_3(x)$ systems are identical.

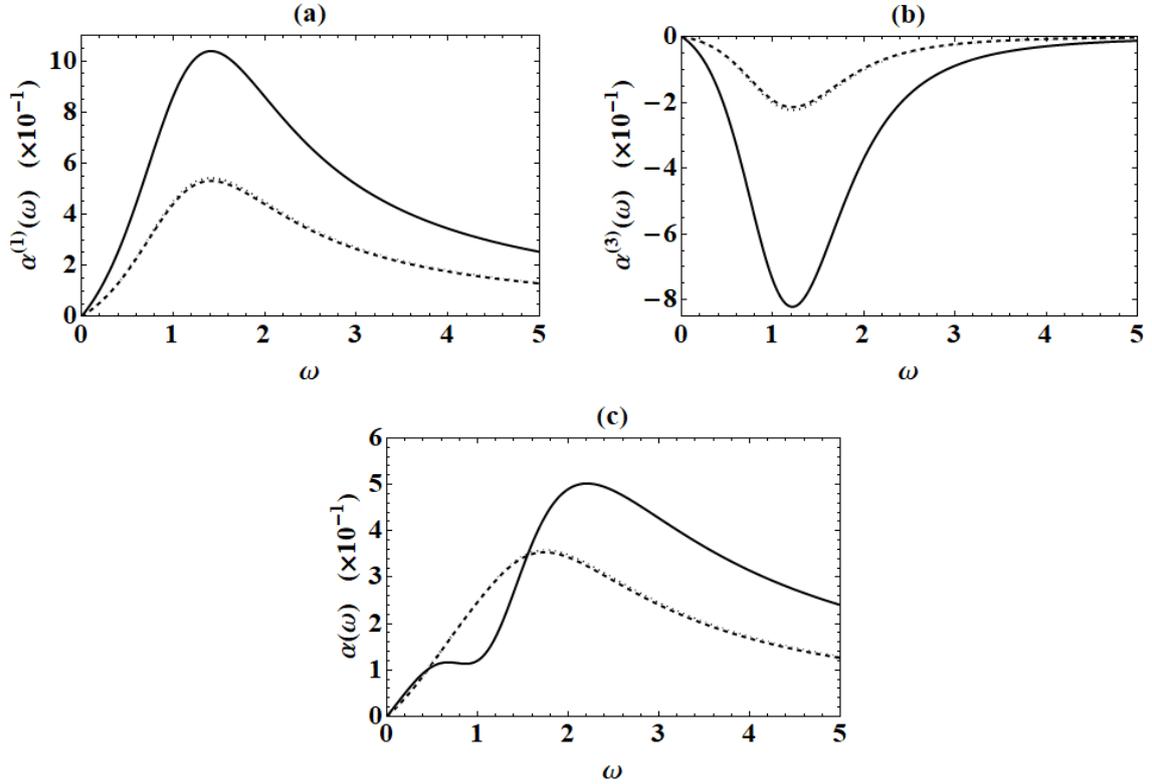

**Fig 1.** Plots of (a) $\alpha^{(1)}(\omega)$, (b) $\alpha^{(3)}(\omega)$ and (c) $\alpha(\omega)$ for $m_1(x)$ (solid line), $m_2(x)$ (dashed line) and $m_3(x)$ (dotted line) systems. In this figure we used $\lambda = 0.6$, $I = 0.5$ and $\mu = \Gamma_0 = \sigma_v = \varepsilon_0 = \varepsilon_R = n_r = c = m_0 = 1$.

We explain the results by analyzing the energies and potential functions of each PDMO system. From Eq. (2.5), we observe that the PDMO systems are a class of isospectral Hamiltonians, *i. e.*, for any mass distribution, the Hamiltonian will be different (Eq. (2.1)), but the energies remain the same, $E_n = n + 1/2$. Consequently, the peak of $\alpha^{(1)}$ remains at $\omega = 1.4$ for any mass distribution in a PDMO Hamiltonian. The potential function (see Eq. (2.1)) for the system described for $m_1(x)$, $m_2(x)$ and $m_3(x)$, are, respectively, given by

$$V_1(x) = \frac{m_0}{2}\left[\frac{\arctan(\lambda x)}{\lambda}\right]^2, \tag{3.4}$$

$$V_2(x) = \frac{m_0}{2}\left[\frac{x}{2}\sqrt{1+(\lambda x)^2} + \frac{\operatorname{arcsinh}(\lambda x)}{2\lambda}\right]^2, \tag{3.5}$$

$$V_3(x) = \frac{m_0}{2}\left\{\sqrt{2}\,\operatorname{arcsinh}[\sqrt{2}\sinh(\lambda x)] - \operatorname{arctanh}\left[\frac{\sinh(\lambda x)}{\sqrt{\cosh(2\lambda x)}}\right]\right\}^2$$

$$\times \frac{\cosh^2(\lambda x)[1+\tanh^2(\lambda x)]}{\lambda^2 \cosh(2\lambda x)}. \tag{3.6}$$

In Fig. 2 we plot $V_1(x)$, $V_2(x)$ and $V_3(x)$ for $\lambda = 0.6$. We observe that $V_1(x)$ is wider than $V_2(x)$ and $V_3(x)$. Owing to this fact, $|\psi_{1,n=0}|^2$ ($|\psi_{1,n=1}|^2$) is wider than $|\psi_{2,n=0}|^2$ ($|\psi_{2,n=1}|^2$) and $|\psi_{3,n=0}|^2$ ($|\psi_{3,n=1}|^2$), and consequently the dipole moment matrix element, $M_{10}$, of the $m_1(x)$ system is greater than those of the $m_2(x)$ and $m_3(x)$ ones, which leads to the differences observed in the $\alpha^{(1)}$ spectra. On the other hand, due to the inversion symmetry of the PDMO systems considered, $M_{11} = M_{00}$ and only the first term inside the brackets in Eq. (2.14) contributes to $\alpha^{(3)}(\omega)$. Therefore, we have $\alpha^{(3)}(\omega) \propto -|M_{10}|^4$ which explains the results shown in Fig. 1(b). The identical widening of $V_2(x)$ and $V_3(x)$ explains the same $\alpha^{(1)}$ and $\alpha^{(3)}$ spectra for the $m_2(x)$ and $m_3(x)$ systems. In Fig. 1(c) we plot $\alpha(\omega)$. We observe that the $m_1(x)$ system exhibits peaks at the frequencies $\omega \approx 0.6$ and $\omega \approx 2.2$, while both $m_2(x)$ and $m_3(x)$ systems exhibit peaks at $\omega \approx 1.6$.

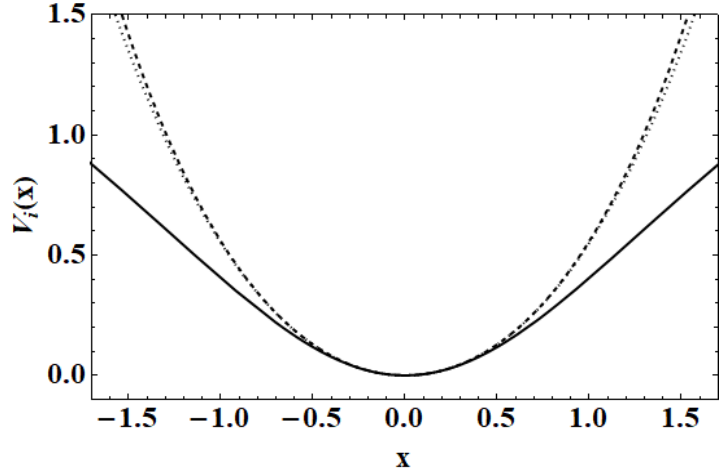

**Fig 2.** Plots of $V_1(x)$ (solid line), $V_2(x)$ (dashed line) and $V_3(x)$ (dotted line). In this figure we used $m_0 = 1$ and $\lambda = 0.6$.

In Figs. 3(a)-(b), we show plots of $\alpha^{(1)}(\omega)$ and $\alpha^{(3)}(\omega)$ for the $m_1(x)$ system using different values of $\lambda$. We observe that $\alpha^{(1)}(\omega)$ $\left(\alpha^{(3)}(\omega)\right)$ increases (decreases) with increasing $\lambda$. Since the denominator in Eq. (2.20) is $\lambda$−independent, the increase (decrease) of $\alpha^{(1)}(\omega)$ $\left(\alpha^{(3)}(\omega)\right)$ is due to $M_{10}$. To explain the variation of $M_{10}$ with $\lambda$, in Fig. 4 (a) we show plots of $V_1(x)$ for some values of $\lambda$. Observe that the higher $\lambda$ the wider $V_1(x)$, leading to the spreading of $|\psi_{1,n=0}|^2$ and $|\psi_{1,n=1}|^2$. This increases the overlapping between the wavefunctions and, consequently, the value of $M_{10}$ which leads to the increase (decrease) of $\alpha^{(1)}(\omega)$ $\left(\alpha^{(3)}(\omega)\right)$. In Figs. 4(b) and 4(c) we plot $V_2(x)$ and $V_3(x)$, respectivelly, for some values of $\lambda$. We observe that $V_2(x)$ and $V_3(x)$ narrow with increasing $\lambda$ which allow us to infer that $\alpha^{(1)}(\omega)$ $\left(\alpha^{(3)}(\omega)\right)$ decreases (increases) for both $m_2(x)$ and $m_3(x)$ systems, as shown in Figs. 3(c)-3(f).

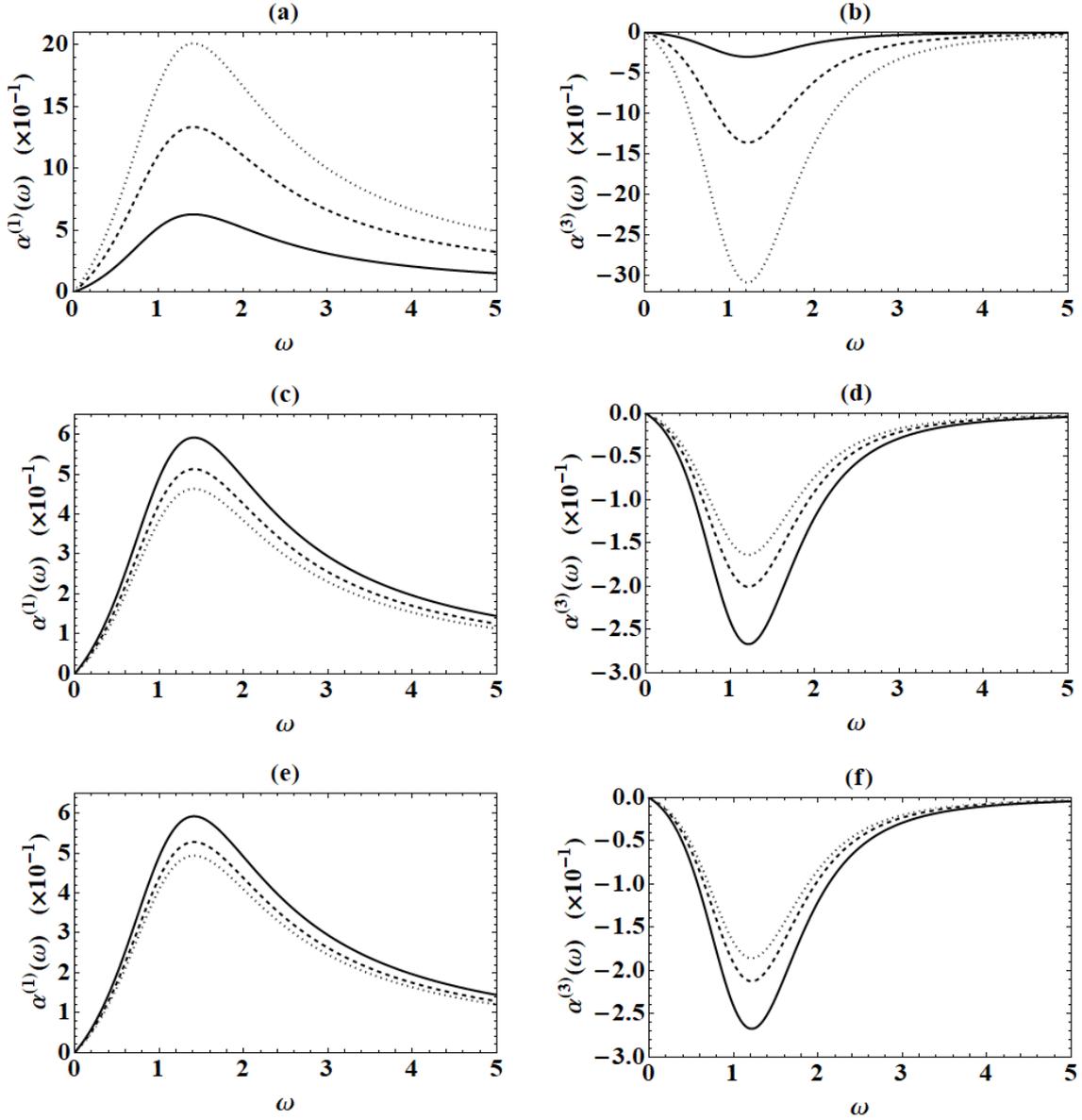

**Fig 3.** Plots of $\alpha^{(1)}(\omega)$ and $\alpha^{(3)}(\omega)$ with $\lambda = 0.2$ (solid line), $\lambda = 0.7$ (dashed line) and $\lambda = 1$ (dotted line), for $m_1(x)$ ((a) and (b)), $m_2(x)$ ((c) and (d)) and $m_3(x)$ ((e) and (f)). In this figure we used $I = 0.5$ and $\mu = \Gamma_0 = \sigma_v = \varepsilon_0 = \varepsilon_R = n_r = c = m_0 = 1$.

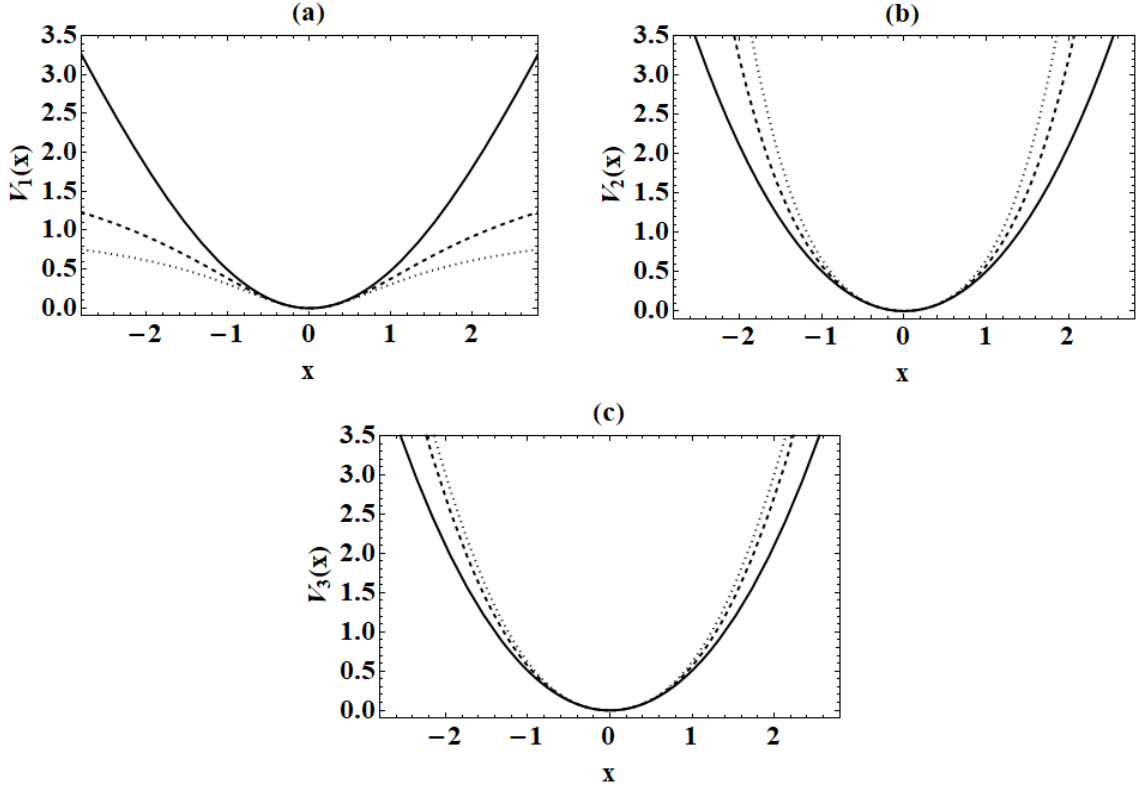

**Fig 4.** Plots of (a) $V_1(x)$, (b) $V_2(x)$ and (c) $V_3(x)$ for $\lambda = 0.2$ (solid line), $\lambda = 0.7$ (dashed line) and $\lambda = 1$ (dotted line). In this figure we used $m_0 = 1$.

In Figs. 5 (a) and (b) we plot $\alpha(\omega)$ with different values of $\lambda$ for $m_1(x)$ and $m_2(x)$, respectively. For $m_1(x)$ we observe that for $\lambda = 0.2$, $\alpha(\omega) > 0$ in the frequency range $\omega \in [0, 5]$, i. e., $\alpha^{(1)}(\omega) > \alpha^{(3)}(\omega)$, and its peak is at $\omega \approx 1.9$. On the other hand, for $\lambda = 0.7$, we observe two regimes: in the frequency range $\omega \notin [0.6, 1.4]$, $\alpha(\omega) > 0$, while in the range $\omega \in [0.6, 1.4]$ the opposite occurs. For this value of $\lambda$, $\alpha(\omega)$ possesses a minimum at $\omega \approx 1.2$. As $\lambda$ increases from $\lambda = 0.2$ to $\lambda = 0.7$, we observe that the peak of $\alpha(\omega)$ suffers a blueshift to $\omega = 2.3$. The increasing of $\lambda$ from $\lambda = 0.7$ to $\lambda = 1$ increases the frequency range for which $\alpha^{(3)}(\omega) > \alpha^{(1)}(\omega)$ ($\omega \in [0, 1.9]$). The minimum value of $\alpha(\omega)$ remains at $\omega \approx 1.2$, however its peak suffers a blueshif to $\omega = 2.9$. From Fig. 5 (b), we observe that $\alpha(\omega)$ increases with the increasing $\lambda$ in the region frequency $\omega \in [1.2, 5]$ and its peak remains at $\omega \approx 1.7$.

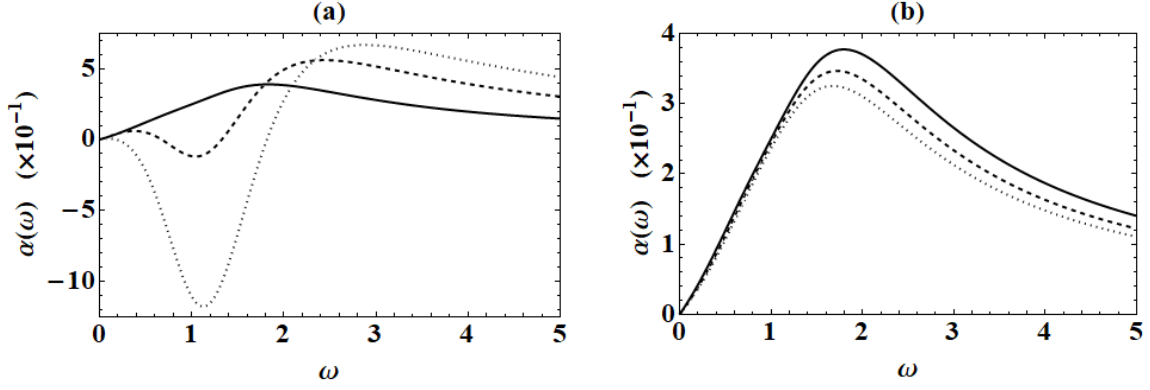

**Fig 5.** Plots of $\alpha(\omega)$ of the PDMO generated by (a) $m_1(x)$ and (b) $m_2(x)$ for $\lambda = 0.2$ (solid line), $\lambda = 0.7$ (dashed line) and $\lambda = 1$ (dotted line). In this figure we used $I = 0.5$ and $\mu = \Gamma_0 = \sigma_v = \varepsilon_0 = \varepsilon_R = n_r = c = m_0 = 1$.

## 4. Concluding remarks

Here we studied the linear $(\alpha^{(1)})$, nonlinear $(\alpha^{(3)})$ and total $(\alpha)$ optical absorptions of position-dependent mass oscillators (PDMOs). We considered three mass distributions used elsewhere [13, 15, 17-19] to describe semiconducting structures, namely: (i) $m_1(x) = \frac{m_0}{[1+(\lambda x)^2]^2}$; (ii) $m_2(x) = m_0[1 + (\lambda x)^2]$, and (iii) $m_3(x) = m_0[1 + \tanh^2(\lambda x)]$.

For all three systems, we observed that the peak (minimum) of $\alpha^{(1)}$ $(\alpha^{(3)})$ always occurs at $= 1.4$ ($\omega = 1.2$), since the systems have the same energy. For a given $\lambda$, we showed that $\alpha^{(1)}$ $(\alpha^{(3)})$ of $m_1(x)$ system is higher (lower) than those of $m_2(x)$ and $m_3(x)$ ones. This is owing to the potential functions of each PDMO system; $V_1(x)$ is wider than $V_2(x)$ and $V_3(x)$. Therefore the wave function of the $m_1(x)$ system are wider than those of the $m_2(x)$ and $m_3(x)$ ones. Consequently, the dipole moment matrix element, $M_{10}$, of the $m_1(x)$ system is greater than those of the other two systems, which leads to the differences observed in the $\alpha^{(1)}(\omega)$ and $\alpha^{(3)}(\omega)$ spectra, since $\alpha^{(3)}(\omega) \propto -|M_{10}|^4$ due to the inversion symmetry. We also observed that in the context of PDMO approach, the $m_2(x)$ and $m_3(x)$ systems can not be distinguished by optical transitions between the two lowest electronic levels.

For the $m_1(x)$ system, we observed that $\alpha^{(1)}(\omega)$ $\left(\alpha^{(3)}(\omega)\right)$ increases (decreases) with increasing $\lambda$. As $E_1 - E_0$ is $\lambda$-independent, this behavior is due to $M_{10}$. The variation of $M_{10}$ with $\lambda$, is due to $V_1(x)$. The higher $\lambda$ the wider $V_1(x)$, which leads to the increasing of the overlapping between the $n = 0$ and $n = 1$ wavefunctions. Consequently, the value of $M_{10}$ increases, which leads to the increase (decrease) of $\alpha^{(1)}(\omega)$ $\left(\alpha^{(3)}(\omega)\right)$. The opposite occurs for the systems $m_2(x)$ and $m_3(x)$, i. e., $V_2(x)$ and $V_3(x)$ narrow with increasing $\lambda$, decreasing of the overlapping between the $n = 0$ and $n = 1$ wavefunctions, which leads to decreasing (increasing) of $M_{10}$ and $\alpha^{(1)}(\omega)$ $\left(\alpha^{(3)}(\omega)\right)$.

## Acknowledgments

The authors are grateful to the National Counsel of Scientific and Technological Development (CNPq) of Brazil for financial support.